\documentclass{emulateapj} \usepackage{apjfonts}

\newcommand{\Ha}{$\rm{H} \alpha$}

\newcommand{\etal}{et~al.}
\newcommand{\NII}{\hbox{[{\rm N}\kern 0.1em{\sc ii}}]}
\newcommand{\HII}{\hbox{{\rm H}\kern 0.1em{\sc ii}}}
\newcommand{\kms}{\hbox{km~s$^{-1}$}}

\newcommand{\ack}{We thank the referee for his/her thorough reading of
  the manuscript. We thank the OWLS team for the use of the
  simulations.  We thank Chris Churchill for useful discussions.  GGK
  was supported by an Australian Research Council Future Fellowship
  FT140100933. KVT and LA acknowledge support by the National Science
  Foundation under Grant \#1410728. Data was obtained at the W.M. Keck
  Observatory, which is operated as a scientific partnership among the
  Caltech, the University of California and the NASA. The Observatory
  was made possible by the generous financial support of the W.M. Keck
  Foundation. Observations were supported by Swinburne Keck programs
  2013B\_W160M and 2014A\_W168M and ANU Keck programs 20132B\_WZ295M
  and 2014A\_Z225M. Part of this work was supported by a NASA Keck PI
  Data Award, administered by the NASA Exoplanet Science
  Institute. The authors wish to recognize and acknowledge the very
  significant cultural role and reverence that the summit of Mauna Kea
  has always had within the indigenous Hawaiian community. We are most
  fortunate to have the opportunity to conduct observations from this
  mountain.}
\slugcomment{resubmitted June 2016} 
\shorttitle{\sc Cold-mode accretion and FMR at $z\sim2$}
\shortauthors{\sc Kacprzak et~al.}

\begin{document}


\title{Cold-mode accretion: Driving the fundamental mass--metallicity relation at $z\sim2$}


\author{\sc Glenn G. Kacprzak\altaffilmark{1}, Freeke van de
  Voort\altaffilmark{2,3}, Karl Glazebrook\altaffilmark{1}, Kim-Vy
  H. Tran\altaffilmark{4}, Tiantian Yuan\altaffilmark{5}, Themiya
  Nanayakkara\altaffilmark{1}, Rebecca J. Allen\altaffilmark{1,7}, Leo
  Alcorn\altaffilmark{4}, Michael Cowley\altaffilmark{6,7}, Ivo
  Labb\'e\altaffilmark{8}, Lee Spitler\altaffilmark{6,7}, Caroline
  Straatman\altaffilmark{8}, Adam Tomczak\altaffilmark{9}}

\altaffiltext{1}{Swinburne University of Technology, Victoria 3122,
Australia {\tt gkacprzak@astro.swin.edu.au}}
\altaffiltext{2}{Department of Astronomy and Theoretical Astrophysics Center, University of California, Berkeley, CA 94720-3411, USA}
\altaffiltext{3}{Academia Sinica Institute of Astronomy and Astrophysics, PO Box 23-141, Taipei 10617, Taiwan}
\altaffiltext{4}{George P. and Cynthia Woods Mitchell Institute for Fundamental Physics and Astronomy, and Department of Physics and Astronomy, Texas A\&M University, College Station, TX 77843-4242, USA}
\altaffiltext{5}{Research School of Astronomy and Astrophysics, The Australian National University, Cotter Road, Weston Creek, ACT 2611, Australia}
\altaffiltext{6}{Department of Physics and Astronomy, Macquarie University, Sydney, NSW 2109, Australia}
\altaffiltext{7}{Australian Astronomical Observatories, PO Box 915 North Ryde NSW 1670, Australia}
\altaffiltext{8}{Leiden Observatory, Leiden University, P.O. Box 9513, 2300 RA Leiden, The Netherlands}
\altaffiltext{9}{Department of Physics, University of California Davis, One Shields Avenue, Davis, CA 95616, USA}
\begin{abstract}
  We investigate the star formation rate (SFR) dependence on the
  stellar mass and gas-phase metallicity relation at $z=2$ with
  MOSFIRE/Keck as part of the ZFIRE survey.  We have identified 117
  galaxies (1.98$\leq z\leq$2.56), with
  $8.9\leq$log(M/M$_{\odot}$)$\leq11.0$, for which we can measure
  gas-phase metallicities. For the first time, we show discernible
  difference between the mass--metallicity relation, using individual
  galaxies, when deviding the sample by low
  ($<10$~M$_{\odot}$yr$^{-1}$) and high ($>10$~M$_{\odot}$yr$^{-1}$)
  SFRs. At fixed mass, low star-forming galaxies tend to have higher
  metallicity than high star-forming galaxies. Using a few basic
  assumptions, we further show that the gas masses and metallicities
  required to produce the fundamental mass--metallicity relation, and
  its intrinsic scatter, are consistent with cold-mode accretion
  predictions obtained from the OWLS hydrodynamical simulations. Our
  results from both simulations and observations are suggestive that
  cold-mode accretion is responsible for the fundamental
  mass--metallicity relation at $z=2$ and demonstrates the direct
  relationship between cosmological accretion and the fundamental
  properties of galaxies.
\end{abstract}




\section{Introduction}
\label{sec:intro}

Mass and metallicity are arguably the most fundamental properties of
galaxies since they reflect stellar build-up/evolution and the cycling
of baryons through outflows and accretion. A clear consequence of
these processes is the galaxy stellar mass--metallicity relation,
which has been observed up to $z\sim4$
\citep[e.g.,][]{tremonti04,erb06,steidel14,troncoso14,zahid14}.

Cosmological simulations continue to show that cosmic accretion
provides significant fuel for star formation resulting in galaxy mass
growth and chemical enrichment via outflows \citep{dekel09,freeke12}.
They also show that most of this gas accretes in the `cold mode',
i.e., without being heated by a virial shock
\citep[e.g.,][]{keres05,fg11,freeke11}. Metal-poor cold-gas accretion has
been observed on the periphery of galaxies, using absorption lines
detected in background quasar spectra
\citep[e.g.,][]{kacprzak12b,crighton13,bouche13}, which are also
dynamically consistent with large-scale cosmic web accretion
\citep[e.g.,][]{steidel02,kacprzak10}.  These inward flows are likely
why, at a given stellar mass, metallicity is dependent on the star
formation rate (SFR) \citep{mannucci10,bothwell13,forbes14,maier15}.

It has been found that at a fixed mass, galaxies with high SFRs have
lower gas-phase metallicities than galaxies with low SFRs. This
fundamental mass--metallicity relation \citep[FMR: see][]{mannucci10}
exhibits little scatter at $z=0$, which implies an equilibrium between
inflowing and outflowing gas and star formation. At higher redshifts,
it is still questionable if a fundamental mass--metallicity relation
exists since at $z\sim2$, the mass--metallicity relation is highly
scattered for stellar masses less than 10$^{10}$M$_{\odot}$
\citep{sanders14,kacprzak15,tran15}.  This observed scatter may not be
surprising given that if a significant metal-poor accretion event
occurs, which is especially common at high redshifts, it is expected
that the galaxy's integrated metallicity decreases and that it
experiences a boost in star formation. Thus, we should expect to see a
fundamental mass--metallicity relation at higher redshifts as well.

The time-scale between accretion and induced star formation is still
uncertain. However, direct observations of metallicity and SFR
profiles of local metal-poor galaxies have shown that highly
star-forming regions within the galaxy have roughly a factor of 10
lower metallicity than the total galaxy metallicity; indicating that
accreted gas is converted into stars within $\sim100$~Myr
\citep{sanchez15}. Thus, this direct relation between accretion, SFR
and metallicity is critical for understanding galaxy formation and
evolution.

We further explore the relationship between SFR and the
mass--metallicity relation at $z\sim2$, and examine the hypothesis
that accretion could cause the scatter in the mass--metallicity
relation.  Here, we present Keck/MOSFIRE observations from the ZFIRE
Survey (Nanayakkara et al.\ submitted) of 117 galaxies for which we
have gas-phase metallicities, stellar masses, SFRs and gas masses
described in \S~\ref{sec:data}. In \S~\ref{sec:results}, we show that
there is a fundamental mass--metallicity relation driven by SFR. We
further use our data, with some underlying basic assumptions, to
compute the gas accretion masses and metallicities, which are shown to
be in agreement with cosmological simulations.  We adopt a $h=0.70$,
$\Omega_{\rm M}=0.3$, $\Omega_{\Lambda}=0.7$ cosmology for our ZFIRE
Survey.
%

\section{MOSFIRE Spectroscopic Observations and Sample}
\label{sec:data}

The ZFIRE survey used Keck/MOSFIRE to spectroscopically identify 181
galaxies to date in the COSMOS field ($1.98\leq z\leq3.26$). The
survey includes a Virgo-like progenitor at $z=2.095$ (57 confirmed
members) with a velocity dispersion of 550~{\kms} \citep{yuan14}. The
spectroscopic targets were selected using the photometric redshifts
from the K-band selected catalog from ZFOURGE \citep{straatman16},
which have an accuracy of $\Delta z/(1+z_{\hbox{spec}})=2\%$
\citep[][; Nanayakkara et al.\ submitted]{yuan14}.

The MOSFIRE near-infrared K--band spectroscopic observations, data
reduction and flux calibration procedures are described in our ZFIRE
Catalog (Nanayakkara et al.\ submitted). All spectra are converted to
vacuum wavelengths.  Our typical $3\sigma$ spectral flux limit is
1.8$\times$10$^{-18}$~ergs/s/cm$^2$.

Gaussian profiles were simultaneously fit to {\Ha} and {\NII}
emission-lines, using both tied line centers and velocity widths, to
determine their total flux.  See \citet{yuan14} and \citet{kacprzak15}
for some example 1D and 2D spectra.  All spectra are flux calibrated
to their total K-band magnitudes to $<$10\% (see Nanayakkara et al.\
submitted).  We compute gas-phase oxygen abundances using the N2
relation of \citet{pettini04} where 12+log(O/H)=8.90+0.57$\times$N2
(N2$\equiv$log(\NII/{\Ha})). We require a $3\sigma$ detection
significance level for {\NII}, otherwise $1\sigma$ detection limits
are used.

Since \citet{kacprzak15} have shown that field and cluster galaxies
exhibit identical mass--metallicity relations at $z=2$ (<0.02dex
offset), we do not differentiate between galaxy environment in this
work. This assumption is further validated by \citet{kewley15} showing
that our $z\sim2$ field and cluster galaxies have consistent ISM
conditions.  After removing AGN \citep[see][]{kacprzak15,cowley16},
our final sample consists of 117 galaxies (1.98$\leq z\leq$2.56,
$<z>=2.160\pm0.008$) with 63 metallicity measurements and 54
metallicity limits. Our spectroscopic sample is mass complete down to
10$^{9.3}$M$_{\odot}$ (see Nanayakkara et al.\ submitted).

We used stellar masses and stellar attenuation ($A_V$) computed from
the photometry from ZFOURGE using \citet{bruzual03} stellar population
models with FAST \citep{kriek09}, assuming solar metallicity, a
\citet{chabrier03} initial mass function, exponentially declining star
formation histories, and constrained to the spectroscopic redshift
\citep{straatman16}.

We derived attenuation corrected {\Ha} SFRs using relations from
\citet{tran15} for dust corrections and from
\citet{hao11} to convert
      {\Ha} fluxes to SFRs (log(SFR)=log($L_{{\rm H}\alpha}$)$-41.27$). Attenuation corrections for both the
      stellar continuum correction \citep{calzetti00} and nebular
      emission \citep{cardelli89} are included \citep[see][for
        details]{tran15}.


\begin{figure*}
\begin{center}
\epsscale{1.15}\plottwo{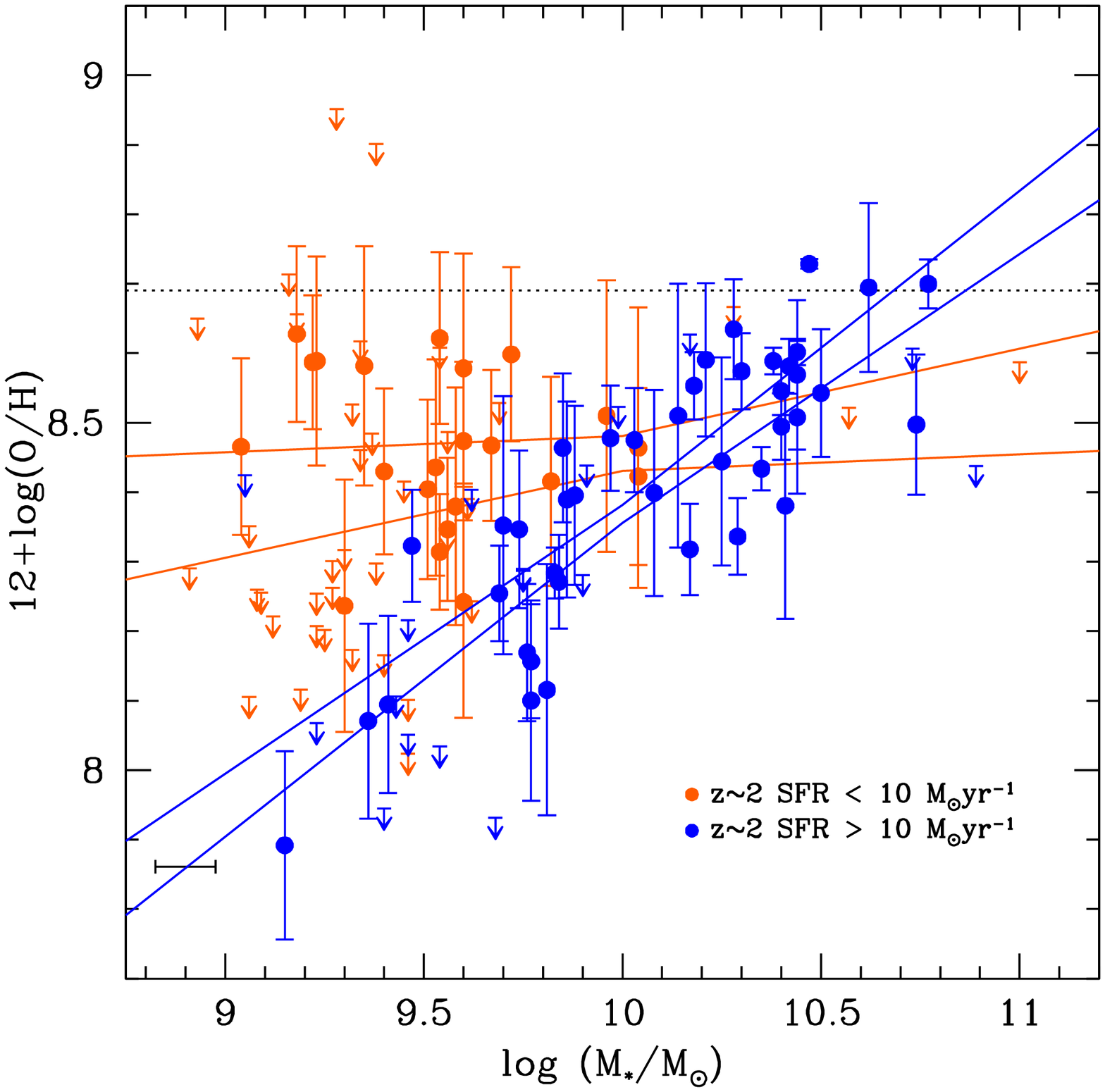}{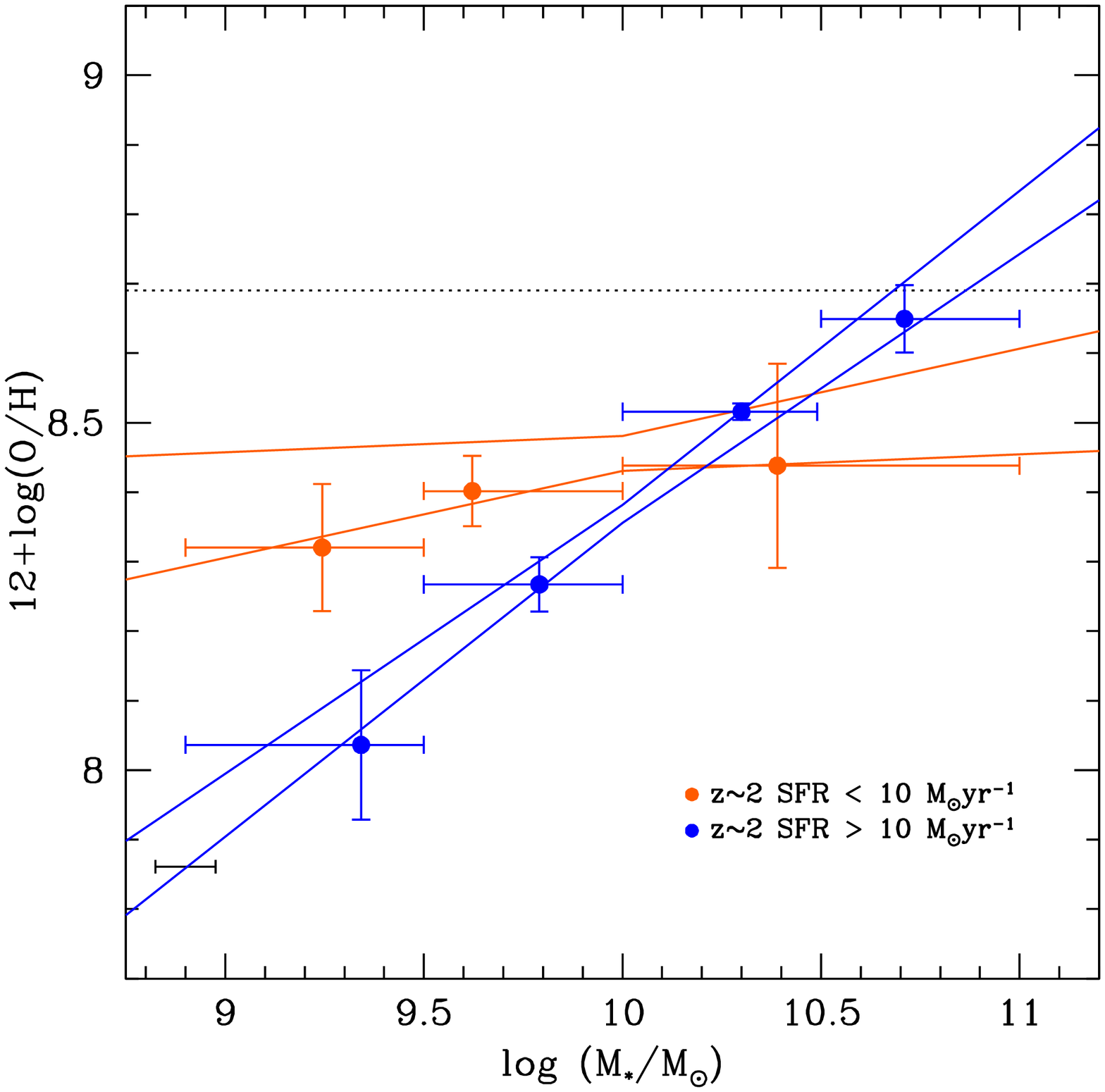}
\caption{The mass--metallicity relation for $z=2$ star-forming
  galaxies. Circles represent metallicity measurements where both
  {\Ha} and {\NII} are detected above a $3\sigma$ significance level
  and downward arrows are quoted as \NII $1\sigma$ detection
  limits. The average error in determining the mass of galaxies using
  ZFOURGE photometry and FAST of $\pm$0.076 is shown in the bottom
  left.  The dotted horizontal line solar abundance
  \citep{asplund09}. The sample is devided into low
  ($<10$~M$_{\odot}$yr$^{-1}$) and high ($>10$~M$_{\odot}$yr$^{-1}$)
  SFR bins. The colored lines are bootstrap fits, with $1\sigma$
  limits, to the mass--metallicity data for low (orange) and high
  (blue) star-forming galaxies. The slopes of the fitted data are
  $m_{_{{\rm SFR}<10}}=0.07\pm0.05$ and $m_{_{{\rm
        SFR}>10}}=0.42\pm0.03$ for the low star-forming
  (SFR$<10$~M$_{\odot}$yr$^{-1}$) and high star-forming
  (SFR$>10$~M$_{\odot}$yr$^{-1}$) galaxies, respectively.--- (right)
  The solid points are stacked spectra within the indicated mass bins
  and note that these are consistent with the fitted data. }
\label{fig:MMR}
\end{center}
\end{figure*}

\section{Results}\label{sec:results}

\subsection{Fundamental Mass--Metallicity Relation}

In Figure~\ref{fig:MMR} (left), we show the mass--metallicity relation
at $z=2$.  We have metallicity measurements for a significant range of
galaxy masses from $8.9\leq$log(M/M$_{\odot}$)$\leq11.0$. As
previously shown in \citet{kacprzak15} our fitted and stacked data are
consistent with previous results from MOSDEF \citep{sanders14},
KBSS-MOSFIRE \citep{steidel14} and \citet{erb06}. We note that,
consistent with previous works, there is a significant amount of
scatter in metallicity for low-mass galaxies compared to what is
observed at $z=0$ \citep[e.g.,][]{tremonti04}.

The mass--metallicity relation in Figure~\ref{fig:MMR} is shown as a
function of low ($<10$~M$_{\odot}$yr$^{-1}$) and high
($>10$~M$_{\odot}$yr$^{-1}$) star forming galaxies. Along with the
data, we present a bootstrap fit (1000 times) using 12+log(O/H)$=y_i
+m_i(M_{*}-10)$.  We fit the data, which include the $1\sigma$ limits,
using the expectation-maximization maximum-likelihood method of
\citet{wolynetz79}. We find for galaxies with
SFR<10~M$_{\odot}$yr$^{-1}$ a $y_{_{{\rm SFR}<10}}=8.46\pm0.03$ with a
flat slope of $m_{_{{\rm SFR}<10}}=0.07\pm0.05$ and for galaxies with
SFR<10~M$_{\odot}$yr$^{-1}$ a $y_{_{{\rm SFR}>10}}=8.37\pm0.01$ with a
steeper slope of $m_{_{{\rm SFR}>10}}=0.42\pm0.03$. The fitted slopes
of the low and high star-forming galaxies differ by $4.4\sigma$.

We further stack the spectra with roughly an equal number of galaxies
per bin. The spectra were stacked, weighting by the uncertainty
spectrum, to determine the typical metallicity per mass bin as shown
in Figure~\ref{fig:MMR} (right). The stacked data are consistent with
the fitted data.  Both the individual metallicity measurements,
stacked data and fits suggest that there is a slope dependence of SFR
on the mass--metallicity relation at the 4.4 sigma level whereby
galaxies with lower SFRs have higher metallicity than galaxies with
higher SFRs for a fixed mass at $M_{*}\lesssim10^{10}$~M$_\odot$.

We have further explored whether high and low star-forming galaxies,
at fixed mass, have morphological differences. We find no discernible
difference in their size and Sersic index distribution between the two
populations.  A visual inspection of the morphologies and nearby
companions of the two populations are also
indistinguishable. Furthermore, both low and high star-forming
galaxies have consistent distributions of stellar attenuation
(${A}_{V}$).



\subsection{Total Gas Mass and Gas Accretion}

%
%

\begin{figure}
\begin{center}
\includegraphics[angle=0,scale=0.45]{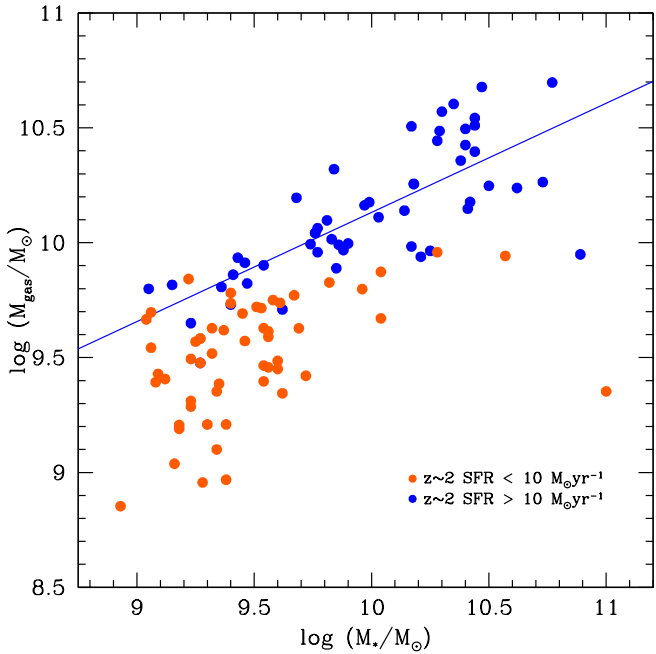}
\caption[angle=0]{The total gas mass derived using the inverse
  Kennicutt-Schmidt Law with the adopted the formalism from
  \citet{papovich15}. The data are bifurcated at a SFR of
  10~M$_{\odot}$yr$^{-1}$. A line is fit to the high SFR galaxies,
  which we use to define a gas mass main sequence whereby low SFR
  galaxies are gas deficient relative to this main sequence
  [M$_{gas}=10.0+0.48(M_{*}-10.0)$].}
\label{fig:Mg}
\end{center}
\end{figure}



\begin{figure}
\begin{center}
\includegraphics[angle=0,scale=1.0]{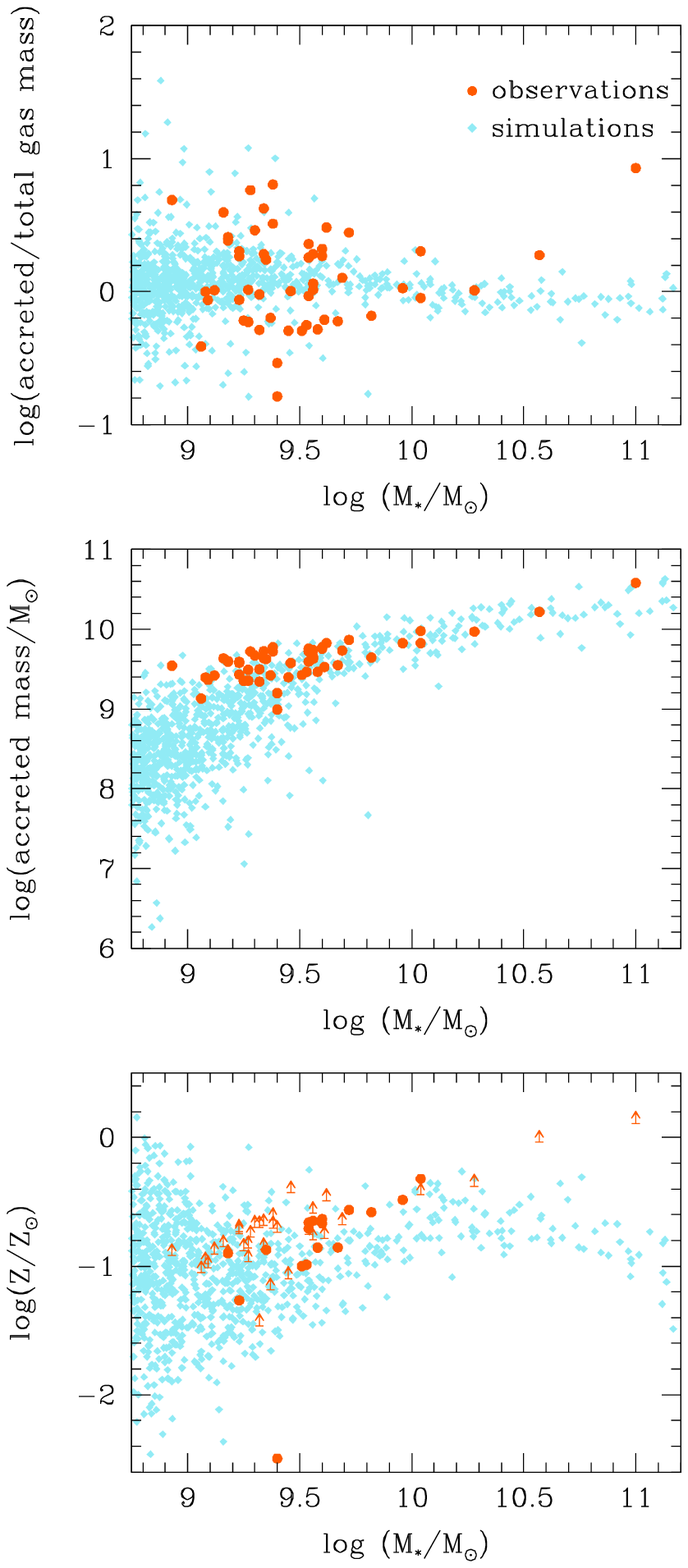}
\caption[angle=0]{(Top) The computed gas fraction of expected
  cold-mode accreted gas mass over the total current gas mass for
  star-forming galaxies as a function of stellar mass from
  observations (orange). The computed cold-mode accreted gas fraction
  from the OWLS simulations over the redshift interval of $z=2-2.5$
  (blue). --- (Middle) The computed total cold-mode accreted gas mass
  as a function of stellar mass from observations (orange) and
  simulations (blue) --- (Bottom) The metallicity of the cold-mode
  accreted gas mass (shown above) from observations (orange) and from
  the OWLS simulations over the redshift interval of $z=2-2.5$
  (blue).}
\label{fig:inflow}
\end{center}
\end{figure}


Although we do not have direct measurements of galaxy gas masses, we
compute galaxy total gas masses using the inverse Kennicutt-Schmidt
Law, adopting the formalism from \citet{papovich15}. Galaxy sizes are
obtained from the catalog of \citet{vdw12}, which were measured from
{\it HST}/WFC3 F160W ($\lambda_{rest}=4860$~\AA) CANDELS multi-cycle
treasury program \citep{grogin11,koekemoer11}.  We obtain 108/117
(92\%) matches within a 0.7$''$ aperture between our sample and the
\citet{vdw12} galaxy size catalog and present total gas masses for
those 108 objects. In Figure~\ref{fig:Mg}, we show the total gas mass
distribution, computed using the inverse Kennicutt-Schmidt Law, as a
function of stellar mass for low and high star-forming galaxies.

Here we test the hypothesis that cosmic gas accretion, which is
predicted to be significant at $z=2$, drives the large scatter in
the mass--metallicity relation and the SFR dependence.  If we simply
assume that galaxies having high SFRs are accreting significant
amounts of near-to or pristine gas, then this would result in a lower
gas-phase metallicity (as seen in Figure~\ref{fig:MMR}). If, however,
the accretion is low or subsiding, for a fixed mass, the galaxy would
have a higher gas-phase metallicity.

In Figure~\ref{fig:Mg}, we fit the high star-forming total gas mass as
a function of stellar mass as the baseline/main sequence of the total
gas expected for an actively accreting galaxy
[M$_{gas}=10.0+0.48(M_{*}-10.0)$]. We assume that the low star-forming
galaxies are gas deficient by a known amount, defined by its offset
from the fit to the high star-forming main sequence. 

We then ask, of the gas that could be added to make deficient galaxies
(orange) become a main sequence object (blue), what would its
metallicity have to be such that it would reside on the
mass--metallicity fitted relation defined by the highly star-forming
galaxies (as shown in Figure~\ref{fig:MMR}).  More simply, the
accretion gas mass is computed from the vertical offset of gas
deficient galaxies (orange) from the main sequence fit in
Figure~\ref{fig:Mg}.  The metallicity of the accreted gas is
determined from the vertical offset of the low star-forming galaxies
(orange) from the best fit of the high star-forming galaxies in
Figure~\ref{fig:MMR}.  Our results are insensitive to the exact slope
of our fit.  Under these basic assumptions, we can determine the mass
and metallicity of gas being accreted in actively accreting galaxies
at $z=2$. We assume an oxygen solar abundance of 8.69
\citep{asplund09} and $Z$ defined as the ratio of the mass of oxygen
in the gas-phase and the hydrogen gas mass. We compute the accretion
gas mass and metallicity as follows:

\begin{equation}\label{eq1}
\Delta Z\Delta M = Z_f(M_o+\Delta M)-Z_oM_o
\end{equation}

\noindent where $Z_o$ is the original metallicity of the metal-rich
low star-forming galaxies (orange points in Figure~\ref{fig:MMR}),
$Z_f$ is the metallicity main sequence defined by the fit to the high
star-forming galaxies shown in Figure~\ref{fig:MMR}. $M_o$ is the
original gas mass of the gas deficient low star-forming galaxies
(orange points in Figure~\ref{fig:Mg}) and $\Delta M$ is accreted gas
added to $M_o$ such that it contains a total gas mass consistent with
the defined gas main sequence in Figure~\ref{fig:Mg}. Finally, $\Delta
Z$ is the metallicity of the gas added ($\Delta M$) required to place
the low star-forming galaxies onto the mass--metallicity relation
defined by the high star-forming galaxies as shown in
Figure~\ref{fig:MMR}. 

In Figure~\ref{fig:inflow} (top) we show the fraction of gas required
to be added/accreted to the gas-deficient low star-forming galaxies to
have a total gas mass that is equivalent to those defined by the high
star-forming galaxy main sequence ($\Delta M$/$M_o$). The majority of
the objects need to acquire/accrete about the same as their existing
gas mass and some objects as much as 8 times more. This amount of gas
required is independent of galaxy stellar mass, however, there is more
scatter at low stellar masses. This is expected since accretion events
are likely to have the largest effect on a galaxy's total mass when it
is low, which is shown in the middle panel of
Figure~\ref{fig:inflow}. Here, accreted gas mass increase with stellar
mass.

Figure~\ref{fig:inflow} (bottom) shows the metallicity required for
all of the low star-forming galaxies to fall on the high star-forming
galaxy mass--metallicity sequence (see Figure~\ref{fig:Mg}, blue
fit). There is a trend with galaxy stellar mass, whereby less massive
galaxies must accrete lower metallicity gas than higher stellar mass
galaxies. The accreted gas has a metallicity range spanning from
around $-1.5$~dex to $-0.25$~dex from low to high stellar mass,
respectively.

\subsection{Cosmological Simulations}


We compare our observational results to cosmological, hydrodynamical
simulations from the OWLS project \citep{Schaye2010}, which uses a
modified version of \textsc{gadget-3} \citep{springel05}. We make use
of the `reference' model with cosmological parameters
$\Omega_\mathrm{m}=0.238$, $\Omega_\mathrm{b}=0.0418$, $h=0.73$,
$\sigma_8=0.74$, $n=0.951$.

The simulation consists of a (25~comoving Mpc~$h^{-1}$)$^3$ periodic
volume with $512^2$ dark matter and $512^2$ baryonic particles with
initial masses $8.7\times10^6$ and $1.9\times10^6$~M$_\odot$,
respectively.  Star formation takes place in gas with density
$n_H\ge0.1$~cm$^{-3}$ and is modeled according to the recipe of
\citet{Schaye2008}. Stellar feedback from Type~{\rm II} supernovae is
implemented kinetically, following \citet{Vecchia2008}, with an
initial wind velocity of 600~\kms and mass loading of 2, which means
that a newly formed star particle kicks twice its own mass in
neighboring gas particles, on average. The abundances of 11 elements
released by massive and intermediate mass stars are followed as
described in \citet{Wiersma2009b}. At $z=2$, there are 665 massive
galaxies with stellar mass above $10^{8.9}$~M$_\odot$. We have
repeated our analysis for a simulation with 8 times lower mass
resolution and found our results to be unchanged.

Galaxy stellar mass is measured within a 20~kpc radius centered on the
galaxy, including only gravitationally bound particles (using
\textsc{subfind}; \citealt{dolag09}). The ISM mass is measured within
the same radius, but with the additional requirement that the gas is
star-forming ($n_\mathrm{H}\ge0.1$~cm$^{-3}$).

We calculated the accreted gas mass from $z=2.5$ to $z=2$, which
corresponds to a time interval of 0.70~Gyr, because we found this
redshift interval to match the normalization of the observationally
derived accreted gas masses. To exclude hot-mode accretion we only
consider gas which has never reached a temperature above $10^{5.5}$~K
before $z=2.5$ \citep{freeke11}.  The accreted gas mass includes the
gas that was in the ISM of the galaxy of interest at $z=2$ and the gas
that was ejected from the ISM between $z=2.5$ and $z=2$, but was not
in the ISM of any galaxy at $z=2.5$ (thus excluding mergers).  The
accreted gas mass includes the gas particles in the ISM of the galaxy
of interest at $z=2$ and the star particles formed after $z=2.5$ and
the gas particles that were ejected from the ISM between $z=2.5$ and
$z=2$, which were not in the ISM of any galaxy at $z=2.5$ (thus
excluding mergers). In order to exclude winds from satellites, we
include only ejected gas that is closer to the galaxy of interest than
any other galaxy.

Figure~\ref{fig:inflow} (top) shows the ratio of the accreted gas mass
from $z=2.5-2$ and the ISM mass at $z=2$ (blue) along with the data
(orange).  Over a period of 0.70~Gyr, the galaxy accretes gas masses
$\sim1-2$ times greater than its total gas mass with some scatter to
higher and lower values for the lower stellar mass galaxies, which is
consistent with our observations.  The total gas mass that is accreted
is shown in Figure~\ref{fig:inflow} (middle). The data exhibit a
similar distribution compared to the simulations, which is stellar
mass dependent. We note that in the simulations, gas originating from
hot-mode accretion has total masses that are 1--1.5~dex lower than gas
masses originating from cold accretion and are inconsistent with our
estimated gas masses from our observations.

The bottom panel of Figure~\ref{fig:inflow} shows the metallicity
distribution of accreted gas from $z=2.5-2$ (blue) along with the
observations (orange). The mean metallicity is calculated from the
particle metallicities at $z=2.5$, before the gas accretes onto the
galaxy. The gas metallicity is dependent on galaxy stellar mass
increasing from low mass to 10$^{10}$~M$_\odot$, which is likely a
direct effect of more massive galaxies enriching their surroundings
more quickly compared to less massive galaxies.  The trend further
decreases at higher masses, which is likely a numerical effect, due to
the fact that the 600~\kms stellar winds are unable to drive
significant galactic outflows. This reduces the ratio of recycled gas
accretion to accretion from the IGM and thus reduces the mean
metallicity. With more efficient feedback at high stellar masses, we
expect the trend of increasing metallicity to continue. The
simulations and observations are consistent and exhibit similar
trends, showing that gas accretion could be responsible for the
scatter in the mass--metallicity relation.

\section{Discussion}\label{sec:conclusion}

We have identified a fundamental mass--metallicity relation at
$z\sim2$ ($4.4\sigma$), whereby galaxies with low star formation rates
($<10$~M$_{\odot}$yr$^{-1}$) exhibit higher metallicities than high
star-forming ($<10$~M$_{\odot}$yr$^{-1}$).

We have examined whether the activity of cold accretion could drive
this scatter in the fundamental mass--metallicity relation.  Given the
simplicity of our assumptions, it is interesting to see that the
metallicities and accreted masses required to reduce the scatter in
the mass--metallicity relation at $z\sim2$ are consistent with and
show similar trends as in our simulations. Therefore, it is tempting
to suggest that gas accretion is solely responsible for the existence
of the fundamental mass--metallicity relation. Our conclusions are
consistent with previous works at low redshift that attribute the
observed scatter in the mass--metallicity relation seen at very low
stellar masses to accretion and/or mergers
\citep{bothwell13,forbes14}. These results demonstrate the direct
relationship between cosmological accretion and the fundamental
properties of galaxies.


We note that at a fixed metallicity, a higher ionization parameter
produces lower {\NII}/{\Ha} ratios and thus lower N2-metallicities
\citep[e.g.,][]{kewley02}. The existence of shock will elevate
{\NII}/{\Ha} ratios \citep[e.g.,][]{rich11,yuan12}.  However, we do
not think ionization parameters/shocks are the main driver for the
scatter of the MZ relation because it is difficult to explain the
bifurcation in SFR since ionization parameter is usually positively
correlated with specific SFR.  We will study the effect of ionization
parameter and/or shocks in future work.

Our interpretation that accretion drives the fundamental
mass--metallicity relation may not be unexpected given that $z\sim2$
is near the peak epoch of star formation, where outflows are
ubiquitous \citep[e.g.,][]{steidel10} and cold accretion likely occurs
\citep[e.g.,][]{bouche13} delivering a significant amount of
metal-poor gas to galaxies. These objects are ideal targets for ALMA
to determine whether high star-forming galaxies indeed have
significantly higher gas fractions and vice versa.

\acknowledgments  

\ack





{\it Facilities:} \facility{Keck I (MOSFIRE)}.

\end{document}